\documentclass[notitlepage,
aps,
twocolumn,
prd,
10pt,
preprintnumbers,
tightenlines,
superscriptaddress,
longbibliography,
nofootinbib]{revtex4-1}
\usepackage[utf8]{inputenc}
\usepackage{amssymb,latexsym}
\usepackage{amsmath,amsbsy,bbm}
\usepackage{epsfig}
\usepackage{bm}
\usepackage{graphicx,comment,color}
\usepackage{slashed}
\usepackage{array}

\usepackage{orcidlink}
\usepackage{graphicx}
\usepackage{dcolumn}
\usepackage{bm}
\usepackage{hyperref}
\hypersetup{colorlinks, urlcolor=blue, linkcolor=blue, citecolor=blue}

\usepackage{tikz}
\usetikzlibrary{quantikz}
\usepackage{algpseudocode}
\usepackage{algorithm}

\begin{document}
\title{Variational Quantum Self-Organizing Map}

\author{Amol Deshmukh\orcidlink{0000-0001-8591-7085}}
\email{amol.deshmukh@ibm.com}
\affiliation{
IBM Quantum, 
IBM T.J. Watson Research Center, 
Yorktown Heights, 
NY, 
USA
}

\date{\today}

\begin{abstract}
We propose a novel quantum neural network architecture for unsupervised learning of classical and quantum data based on the kernelized version of Kohonen's self-organizing map.
The central idea behind our algorithm is to replace the Euclidean distance metric with the fidelity between quantum states to identify the best matching unit from the low-dimensional grid of output neurons in the self-organizing map. 
The fidelities between the unknown quantum state and the quantum states containing the variational parameters are estimated by computing the transition probability on a quantum computer. 
The estimated fidelities are in turn used to adjust the variational parameters of the output neurons.
Unlike $\mathcal{O}(N^{2})$ circuit evaluations needed in quantum kernel estimation, our algorithm requires $\mathcal{O}(N)$ circuit evaluations for $N$ data samples.
Analogous to the classical version of the self-organizing map, our algorithm learns a mapping from a high-dimensional Hilbert space to a low-dimensional grid of lattice points while preserving the underlying topology of the Hilbert space. 
We showcase the effectiveness of our algorithm by constructing a two-dimensional visualization that accurately differentiates between the three distinct species of flowers in Fisher's Iris dataset. 
In addition, we demonstrate the efficacy of our approach on quantum data by creating a two-dimensional map that preserves the topology of the state space in the Schwinger model and distinguishes between the two separate phases of the model at $\theta = \pi$.
\end{abstract}

\maketitle

\section{\label{sec:intro}Introduction}
Investigations into the inner workings of the biological neural networks continue to inspire novel, non-von Neumann architecture-based techniques of computation. 
Artificial neural networks are the prime and very successful examples of such techniques that have revolutionized the way large datasets are processed for problems in pattern recognition and machine learning~\cite{Goodfellow-et-al-2016, SCHMIDHUBER201585}. 
Artificial neural networks have demonstrated significant utility across a wide range of scientific disciplines, such as neuroscience, physics, material science, logistics, and finance, among others~\cite{YANG20201048, 1999966, su12093760, huang2020deep, heaton2017deep}.

Inspired by the competitive learning rules of the locally ordered neurons in the cerebral cortex responsible for the `associative memory' formation in the brain~\cite{LEVY1979233, bliss1993synaptic, mcclelland1995there, schmidt2019self}; and in an attempt to capture and formalize their learning abilities, Kohonen introduced the idea of \textit{self-organizing map} (SOM)~\cite{kohonen1982self, 58325}.
SOM can be conceived as a two-layered feedforward neural network with all-to-all connections from the input layer to the neurons in the output layer. 
It has been extensively studied and used as a tool for unsupervised machine learning tasks, such as clustering, dimensionality reduction, and for constructing low-dimensional topology-preserving visualizations of high-dimensional datasets.
SOMs have been very successfully applied for a variety of use cases in finance, engineering, natural language processing, sociology, etc.~\cite{537105, deboeck2013visual, KALTEH2008835}. 

Typically, training SOMs on large datasets is a time-consuming and costly process.
Moreover, if the data has a sufficiently complicated structure, SOMs may be unable to uncover the hidden structure within those datasets.
Utilizing efficient algorithms or hardware accelerators, such as GPUs, to expedite the training of SOMs has been the subject of numerous research studies~\cite{McConnell_2012, JSSv078i09, lawrence1999scalable}. 
How does new hardware, such as a quantum processor, improve or alter the nature of this machine learning algorithm, is the next question that naturally arises. 

Quantum machine learning techniques which use quantum processors, instead of a classical (either a CPU-based or a GPU-based) computer have recently risen into prominence.
Quantum machine learning refers to the emergent field of study that analyzes the applications of quantum algorithms for machine learning tasks on either classical or quantum data~\cite{biamonte2017quantum, cerezo2022challenges}.
The holy grail of this field is to find algorithms that can provide a \textit{quantum advantage} over classical machine learning techniques on either classical or quantum data~\cite{liu2021rigorous, PhysRevLett.113.130503, huang2022quantum} (although this is a debated topic, see~\cite{PRXQuantum.3.030101}). 
The term quantum advantage can encompass multiple aspects, including the speed-up achieved by the quantum algorithms (time complexity), the sample requirements for training and inference (sample complexity), or the performance metrics of the models~\cite{anshu2023survey, havlivcek2019supervised, PhysRevLett.122.040504, mensa2023quantum}.

A plethora of quantum machine learning algorithms has been created for supervised machine learning tasks, including classification, regression, and feature selection, encompassing both fault-tolerant devices as well as near-term quantum computers~\cite{PERALGARCIA2024100619, zhang2020recent, du2025quantum}. 
Numerous approaches for unsupervised learning have also been suggested; however, they tend to be either fault-tolerant or linear in nature~\cite{lloyd2013quantum, aimeur2013quantum, lloyd2014quantum, NEURIPS2019_16026d60}, 
though certain studies have addressed both limitations~\cite{PhysRevResearch.4.043199}.
The current constraints on the number of qubits and the occurrence of gate errors further confine us to hybrid quantum-classical machine learning methodologies.
To address these deficiencies in the literature, this work presents a novel unsupervised quantum machine learning technique that draws inspiration from Kohonen's \textit{self-organizing map} concept.
We utilize principles from quantum kernel approaches to identify the best matching unit on quantum computers. 
The design of our quantum neural network is directly motivated by the architectural framework of the classical self-organizing map (SOM).

Additionally, we are motivated by another relevant question concerning the representation of quantum states \cite{PhysRevLett.105.150401, carrasquilla2019reconstructing, torlai2018neural}. 
The pertinent issue is the depiction of states produced by quantum computers. 
Is there a more efficient method for visualizing quantum states with exceedingly large dimensions?
One may contemplate augmenting the method to discern the topological structure intrinsic to quantum data, which may be produced from the simulations of quantum many-body systems. 
An unsupervised machine learning algorithm may assist in identifying distinct phases of such systems~\cite{PhysRevA.105.042432, PhysRevB.94.195105, PhysRevE.96.022140, PhysRevResearch.3.013074, wang2020nuclear, ni2019machine}. 
A low-dimensional map can be created that may correspond to different phases of these many-body systems.
A similar study has already been performed with classical self-organizing maps~\cite{PhysRevB.99.041108}. 
In this context, we apply our method to the Schwinger model, a simplified representation of quantum electrodynamic interactions in $1+1$-dimensional spacetime.

The paper is organized as follows. 
In section \ref{sec:background}, we provide the necessary background information and outline our framework.
We introduce the original version of Kohonen's self-organizing map, along with its modified version designed to operate in kernel space.
This section will also establish the notations used throughout the work.
In section \ref{sec:quantum_som}, we introduce the proposed algorithm: the variational quantum self-organizing map or \textit{Variational QSOM}. In section \ref{sec:experiments}, we demonstrate the effectiveness of our algorithm on a classical data set (Fisher's Iris data). 
Additionally, the phase space of the Schwinger model at $\theta=\pi$ is analyzed using the proposed algorithm. 
Finally, in section \ref{sec:conclusion}, we discuss the theoretical aspects such as runtime complexity, robustness, and possible extensions of our algorithm.

\section{\label{sec:background}Background}
In order to understand self-organizing maps (SOMs), it is helpful to provide a concise overview of the basic concepts of artificial neural networks as a whole, and then explain the unique features of SOMs within this context. 
There are three main types of biologically inspired artificial neural networks which are primarily studied in the deep learning literature: \textit{signal-transfer networks}, \textit{state-transfer networks}, and \textit{competitive-learning networks}~\cite{58325}. 
Layered feed-forward neural networks such as multilayer perceptrons and convolutional neural networks are some of the well-known examples of the signal-transfer networks. 
In these types of neural networks, the output signal has a unique dependence on the input signal. 
Learning in these networks is governed by the error-correcting back-propagation algorithms. 
State-transfer networks, on the other hand, are based on relaxation effects. 
The strong nature of the feedbacks and non-linearities in these networks drives the internal states of the neurons to stable fixed points in the phase space. 
Hopfield networks and Boltzmann machines are exemplary networks from this category. 
The learning of these networks is governed by Hebbian learning or Boltzmann learning rules~\cite{hebb1949organization, ackley1985learning}.
As the name suggests, the learning of the networks in the third category is governed by competitive learning. 
The neurons in the output layer receive identical input information from the input layer, and they compete in their activities. 
Through lateral interactions, one of the neurons becomes the `winner' and suppresses the activities of all the other neurons in the output layer.
The winner neurons alternate depending on the received input. 
Once trained, different sections of the networks become sensitive to different parts of the vectorial input signals.

Based on such competitive learning rules, Kohonen introduced the idea of self-organizing maps inspired by the biological semantic or topographic maps created in the mammalian brain. 
SOMs have been used as a tool for unsupervised ML tasks, such as clustering, dimensionality reduction, and for learning a low-dimensional topology-preserving representation of the high-dimensional data sets.

\subsection{\label{sec:linear_som}Kohonen's self-organizing map}
Kohonen's map or self-organizing map (SOM) is a type of competitive-learning network which is widely used for unsupervised learning.
SOM is composed of two layers, an input layer and an output layer. 
The input layer feeds a data sample from the high-dimensional dataset $\bm{x}\in \Omega$, whereas the output layer consists of a lattice of neurons $\{l_1,\dots l_k\}\in L$ arranged in a particular topology, for example, a rectangular or hexagonal grid.
To each lattice point in the grid is associated a weight vector $\bm{w}_i\in \Omega$.
The main objective of the self-organizing map is to learn the topology-preserving map $\kappa(\bm{x})$ from a high dimensional continuous space $\Omega$ (usually $\mathbb{R}^N$, where $N$ corresponds to the dimensionality of $\bm{x}$) to a low-dimensional lattice space $L$ of $k$ neurons 
\begin{equation}
\kappa(\bm{x}): \Omega \to L.
\end{equation}
The learning of this mapping, with the help of data, proceeds in three iterative steps as follows (see Fig.~\ref{fig:classical_som}).
In the first step, a sample $\bm{x}$ is selected randomly from the data.   
The second step comprises of finding the neuron $l^*$ (i.e. $l_{i^*}$), termed \textit{best matching unit (BMU)}, which is closest to $\bm{x}$.
This is achieved by calculating the similarity score between $\bm{x}$ and the weight vector $\bm{w_i}$ associated with each neuron $l_i\in L$
\begin{equation}
\label{eq:bmu_rule}
||\bm{x}-\bm{{w}}_{i^*}^t|| =  \text{min}_i\Big\{||\bm{x}-\bm{w}^t_{i}||\Big\},
\end{equation}
where, superscript $t$ indicates $t^{\text{th}}$ iteration.
Typically one uses Euclidean metric to calculate the similarity score; in principle however other distance metrics can also be used. 
In the final step, only those neurons which lie in the neighbourhood of the winning neuron, defined by $h(l_i,l^*)$ are activated and the weights ${w}_i$ corresponding to those neurons are updated as follows:
\begin{equation}
\label{eq:weight_update_rule}
\bm{w}_i^{t+1} = \bm{w}_i^t + \alpha^t h^t \big(\bm{x}-\bm{w}_i^t\big),
\end{equation}
where, $h^t(l^*, l_i)$ is the neighbourhood function and $\alpha^t$ is the learning rate. 
The explicit dependance of $t$ in both the learning rate $\alpha$ and neighborhood function $h(l^*, l_i)$ reflects the fact that these are iteration dependent quantities and are usually artificially reduced in magnitude at each iteration.
Intuitively, the neurons that are closest to the BMU are updated in the direction of the sample $\bm{x}$, whereas the neurons which are far away from BMU are updated the least. 
Along with the learning rate $\alpha$, the form of the neighborhood function is a hyperparameter of the algorithm, but usually the Gaussian functional form 
\begin{equation}
\label{eq:neighborhood}
h(\bm{d}_{ij}) = \exp\Big(-\frac{\bm{d}_{ij}^2}{2\sigma^2}\Big),
\end{equation}
is preferred. 
Here, $\bm{d}_{ij}$ refers to the distance between the nodes $\bm{l}_i$ and $\bm{l}_j$.
The completion of the training process results in a map $\kappa(\bm{x})$ from the high-dimensional space $\Omega$ to the lattice space of neurons $L$, which preserves the topological structure present in $\Omega$. 
During the inference phase, the weights of the neural network are not updated. Instead, the appropriate BMU $l^*$ is picked for a given data sample $\bm{x}$.

In contrast to the alternative methods of dimensionality reduction, SOM exhibits non-parametric and nonlinear mapping characteristics. The algorithm is robust, the learning is time-efficient, and the outlines of the map are developed quickly~\cite{anderson1995introduction}.
\begin{center}
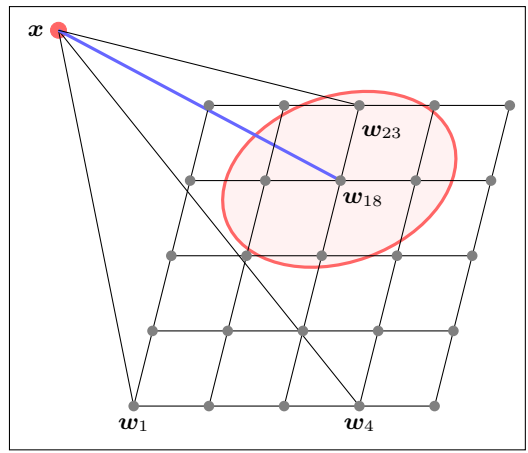
\begin{figure}
\begin{tikzpicture}[framed]
\filldraw [red!60] (-1,5) circle [radius=3pt];
\filldraw[color=red!60, fill=red!5, very thick, rotate=20](2.60+1, 3-1.1) circle (1.6 and 1.1);
\draw(-1,5) -- (0,0);
\draw (-1,5) -- (3,0);
\draw[color=blue!60, very thick] (-1,5) -- (2.75, 3);
\draw (-1,5) -- (3,4);
\draw(-1.3, 5) node{$\bm{x}$};
\foreach \x in {0, 1, 2, 3, 4}
    \draw[black, thin] (\x, 0) -- (\x+1, 4);
\foreach \y in {0, 1, 2, 3, 4}
    \draw[black, thin] (0 + + \y * 0.25, \y) -- (4 + \y * 0.25, \y);
\foreach \x in {0, 1, 2, 3, 4}
    \foreach \y in {0, 1, 2, 3, 4}
        \fill[gray] (\x + \y * 0.25, \y) circle [radius=2pt];
\draw(0.0, -0.25) node{$\bm{w}_1$};
\draw(3, -0.25) node{$\bm{w}_4$};
\draw(2.3 + 0.75, 2.75) node{$\bm{w}_{18}$};
\draw(3.3, 3.65) node{$\bm{w}_{23}$};
\end{tikzpicture}
\caption{A snapshot of the (classical) self-organizing map during the training phase. First, the weights are randomly initialized and then the best matching unit (BMU) is found (here, $l_{18}$) by calculating the Euclidean distance between $\bm{x}$ and all the weights $\{\bm{w}_1, \dots, \bm{w}_k\}$, see Eq.~(\ref{eq:bmu_rule}). The weights in the neighbourhood (here represented by a red circle) of the winning neuron are updated using the update rule specified in Eq.~(\ref{eq:weight_update_rule}).}
\label{fig:classical_som}
\end{figure}
\end{center}

\subsection{\label{sec:kernelized_som}Kernelized self-organizing map}

Kohonen's original method was refined by Andras by integrating scenarios in which the original data vector must be projected into a higher-dimensional feature space in order to accurately identify the best matching unit~\cite{andras2002kernel}.
Kernel methods, a well-known concept in the field of machine learning, are used in this improved algorithm.

Kernel approaches are prevalent in pattern recognition, machine learning, and artificial intelligence~\cite{10.1214/009053607000000677}. 
They are very useful for distinguishing between non-linearly separable classes of data points. 
The fundamental principle behind kernel methods is that the Euclidean inner product between two data vectors (as seen in the preceding algorithm) can be replaced with an equivalent inner product in a higher-dimensional feature space. 
The simple substitution
\begin{equation}
\bm{x}_i.\bm{x}_j \to k(\bm{x}_i, \bm{x}_j)=\phi(\bm{x}_i).\phi(\bm{x}_j),
\end{equation}
essentially allows one to explore the higher dimensional spaces $\phi(\bm{x})$ without explicitly constructing the mapping between the original Euclidean space to the higher dimensional space.
Kernel approaches excel when the inner product of two data vectors can be calculated more efficiently than the calculation of the explicit higher dimensional map.

Andras observed that the updates of the weight vectors do not need to take place in higher dimensional spaces.
The observations made were as follows:
In Kohonen's original implementation, the update rule for the weights in the self-organizing map has an alternative interpretation of the minimization of the $||\bm{x}-\bm{w}_{i}||^2$, which can be seen as:
\begin{equation}
\label{eq:weight_update_rule_reinterpret}
\bm{w}_i^{t+1} = \bm{w}_i^t - \bar{\alpha}^t h^t \frac{\partial}{\partial \bm{w}_i}\big[||\bm{x}-\bm{w}_i^t||^2\big].
\end{equation}
Now, consider the mapping of a data vector $\bm{x}$ to a higher dimensional space $\Psi(\bm{x})\in\Omega$. 
All the weight vectors $\bm{w}_i$'s will also be mapped to the same higher dimensional space $\Psi(\bm{w}_i)\in\Omega$. 
The next step consists of minimizing the distance between the data vector and the weight vectors mapped to the higher dimensional spaces, i.e.
\begin{equation}
\label{eq:kernel_bmu_rule}
||\Psi(\bm{x})-\Psi(\bm{w}^t_{i^*})|| =  \text{min}_i\Big\{||\Psi(\bm{x})-\Psi(\bm{w}^t_{i})||\Big\}.
\end{equation}
The distance between the two higher dimensional vectors can be expressed in terms of the kernel function as 
\begin{align}
||\Psi(\bm{x})-\Psi(\bm{w})||^2 
&= \langle\Psi(\bm{x})-\Psi(\bm{w}), \Psi(\bm{x})-\Psi(\bm{w})\rangle\nonumber\\
&= K(\bm{x}, \bm{x}) + K(\bm{w}, \bm{w}) - 2 K(\bm{x}, \bm{w}),
\end{align}
where, $K(\bm{x}, \bm{y}) = \langle\Psi(\bm{x}), \Psi(\bm{y})\rangle$. Following the gradient descent procedure for the minimization of this distance, we arrive at the following rule for the updates of the weights in the kernelized self-organizing map:
\begin{align}
\label{eq:kernel_weight_update_rule}
\bm{w}_i^{t+1} = \bm{w}_i^t - \alpha^t h^t \Big(\frac{\partial}{\partial \bm{w}_i}K(\bm{w}, \bm{w})\Big|_{\bm{w}_i^t} - 2\frac{\partial}{\partial \bm{w}_i}K(\bm{x}, \bm{w})\Big|_{\bm{w}_i^t}\Big).
\end{align}
The completion of the training process results in a map $\kappa(\bm{x})$ from the high-dimensional space $\Omega$ to $L$
\begin{equation}
\kappa(\Psi(\bm{x})): \mathbb{R}^N \to \Omega \to L,
\end{equation}
which preserves the topological structure present in $\Omega$.
In the inference phase, similar to the original version of the algorithm, the weights are no longer updated; instead, the appropriate BMU for a data sample $x$ is chosen.
It was demonstrated that the kernelized version of the algorithm exhibited superior performance compared to the original version. This can be attributed to its ability to accurately identify the BMU in higher-dimensional spaces~\cite{andras2002kernel}.

\section{\label{sec:quantum_som}Variational Quantum Self-Organizing Maps}
Equipped with the understanding of the classical self-organizing map in section~\ref{sec:linear_som} and its kernelized counterpart in section~\ref{sec:kernelized_som}, we are now ready to discuss the variational quantum self-organizing map (variational QSOM) for the unsupervised learning of the classical and quantum data. 

Similar to the classical self-organizing map, variational QSOM is composed of two layers, an input layer and an output layer.
The input layer feeds a data sample from the quantum dataset $\rho(x)\in \mathcal{B(H)}$, whereas the output layer consists of a lattice of neurons $\{l_1,\dots l_k\}\in L$ arranged in a particular topology, for example, a rectangular or hexagonal grid, as shown in Fig.~\ref{fig:quantum_som}.
$\rho(\bm{x})$ corresponds to either the quantum data sample generated from a quantum computer or a classical data sample mapped to a quantum state via a quantum feature map, given by 
\begin{equation}
    \Phi: \bm{x} \to \rho(\bm{x})= |U(\bm{x})\rangle\langle U^\dagger(\bm{x})|,
\end{equation}
where $U(\bm{x})$ is a unitary composed of quantum gates. 
It should be noted that, to ensure the possibility of a quantum advantage on an unsupervised machine learning task it is necessary (but not sufficient) to establish that the map generated by $U(\bm{x})$ is sufficiently complex to be non-simulable on a classical computer.
Note, $\mathcal{B(H)}$ refers to the set of all bounded operators on the Hilbert space of quantum states.
To each lattice point in the grid is associated a parameterized quantum state $\rho(\bm{\theta}_i)\in \mathcal{B(H)}$.
The $\bm{\theta}_i$'s here, may correspond to the angles in the Pauli-rotation gates.
In analogy with the classical SOM, the main objective of the variational QSOM is to learn the topology-preserving map $\kappa(\rho(\bm{x}))$ from a high dimensional continuous space $\mathcal{B(H)}$ to a low-dimensional lattice space $L$ of $k$ neurons 
\begin{equation}
\kappa(\rho(\bm{x})): \mathbb{R}^N \to \mathcal{B(H)} \to L.
\end{equation}

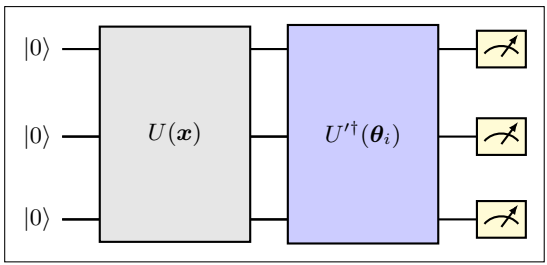
\begin{figure}
\begin{center}
     \tikzset{
         meter/.append style={fill=yellow!20},         
         my label/.append style={above right,xshift=0.3cm},
         phase label/.append style={label position=above}
        }
\begin{quantikz}[framed]
\lstick{$\ket{0}$}& \gate[3, style={fill=gray!20}][2cm]{U(\bm{x})} & \gate[3, style={fill=blue!20}][2cm]{U'^\dagger(\bm{\theta}_i)} & \meter{} \\ 
\lstick{$\ket{0}$}& \qw & \qw & \meter{}\\
\lstick{$\ket{0}$}& \qw & \qw & \meter{}
\end{quantikz}
\caption{Quantum circuit to estimate the transition probability between an unknown quantum state and the quantum state corresponding to a neuron. 
The goal of this architecture is to calculate the degree of overlap with each neuron's weights and identify the one that generates the highest number of `$0\dots0$' bitstrings.}
\label{fig:transition_probability}
\end{center}
\end{figure}

\begin{center}
\begin{figure}
\begin{tikzpicture}[framed]
\filldraw [red!60] (-1,5) circle [radius=3pt];
\filldraw[color=red!60, fill=red!5, very thick, rotate=20](2.60+1, 3-1.1) circle (1.6 and 1.1);
\draw[gray] (-1,5) -- (0,0);
\draw[gray] (-1,5) -- (3,0);
\draw[color=blue!60, very thick] (-1,5) -- (2.75, 3);
\draw[gray] (-1,5) -- (3,4);
\draw(-1.45, 5) node{$\rho(x)$};
\foreach \x in {0, 1, 2, 3, 4}
    \draw[black, thin] (\x, 0) -- (\x+1, 4);
\foreach \y in {0, 1, 2, 3, 4}
    \draw[black, thin] (0 + + \y * 0.25, \y) -- (4 + \y * 0.25, \y);
\foreach \x in {0, 1, 2, 3, 4}
    \foreach \y in {0, 1, 2, 3, 4}
        \fill[gray] (\x + \y * 0.25, \y) circle [radius=2pt];
\draw(0.0, -0.25) node{$\rho({\theta}_{1})$};
\draw(3, -0.25) node{$\rho({\theta}_{4})$};
\draw(2.3 + 0.85, 2.75) node{$\rho({\theta}_{18})$};
\draw(3.4, 3.75) node{$\rho({\theta}_{23})$};
\end{tikzpicture}
\caption{A snapshot of the variational quantum self-organizing map during the training phase. Analogous to the classical self-organizing map, first the weights are randomly initialized and then the best matching unit (BMU) is found (here, $l_{18}$) by calculating the Hilbert-Shmidt norm between $\rho(x)$ and the quantum states corresponding to the assigned weights $\{\rho({{\theta}_1}), \dots, \rho({{\theta}_{k}})\}$, see Eq.~(\ref{eq:quantum_bmu_rule}). The weights in the neighbourhood (here represented by a red circle) of the winning neuron are updated using the update rule as specified in Eq.~(\ref{eq:simple_quantum_weight_update_rule}).}
\label{fig:quantum_som}
\end{figure}
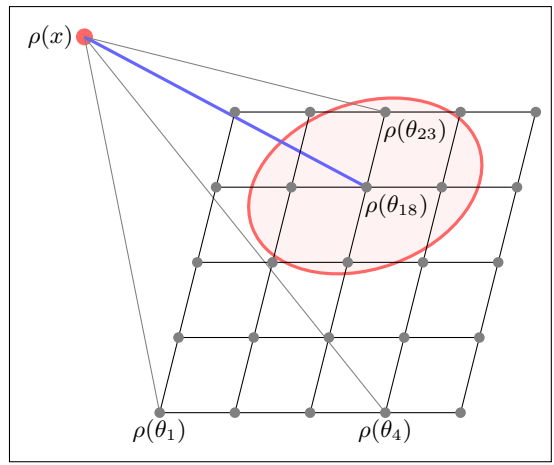
\end{center}
The learning of this mapping, with the help of quantum data, proceeds in three iterative steps as follows.
A sample quantum state $\rho(\bm{x})$ is selected randomly from the given data. 
The second step comprises of finding the neuron $l^*$, i.e. \textit{best matching unit (BMU)}, which is closest to $\rho(\bm{x})$.
\begin{equation}
\label{eq:quantum_bmu_rule}
    d\Big[\rho(\bm{x}), \rho(\bm{\theta}^t_{i^*})\Big] =  \text{min}_i\Big\{d\Big[\rho(\bm{x}), \rho(\bm{\theta}^t_i))\Big]\Big\}.
\end{equation}
This is achieved by calculating the Hilbert-Shmidt inner product between $\rho(\bm{x})$ and the weight vector $\rho(\bm{\theta}_i)$ associated with each neuron $l_i\in L$, as
\begin{equation}
    K(\bm{x}, \bm{\theta}_i) = \text{Tr}\big[\rho(\bm{x})\rho(\bm{\theta}_i)\big] =|\langle0|U^\dagger(\bm{x})U(\bm{\theta}_i)|0\rangle|^2.
\end{equation}
This is a crucial step in our algorithm; where the distance between the two states is estimated on a quantum computer by calculating the transition probability between the two states $\rho(\bm{x})$ and $\rho(\bm{\theta}_i)$, as shown in Fig.~\ref{fig:transition_probability}.
We also note that, for the unsupervised machine learning tasks, the embedding of the data sample and the weight vectors do not need to be identical. 
In theory, this algorithm can be used to calculate the overlap between two different quantum feature maps; thus, $U=U'$ is merely an exception.

In a manner analogous to that of the kernelized self-organized map, the rule that governs the weights (i.e. angles in the parametrized quantum gates) update process is given as
\begin{equation}
\label{eq:quantum_weight_update_rule}
\bm{\theta}_i^{t+1} = \bm{\theta}_i^t - \alpha^t h^t\Big(\frac{\partial}{\partial \bm{\theta}}K(\bm{\theta}, \bm{\theta})\Big|_{\bm{\theta}_i^t} - 2\frac{\partial}{\partial\bm{\theta}}K(\bm{x}, \bm{\theta})\Big|_{\bm{\theta}_i^t}\Big).
\end{equation}
In those cases where $U=U'$, the first term in Eq.~(\ref{eq:quantum_weight_update_rule}) vanishes, since $K(\bm{\theta}, \bm{\theta}) = 1$, and thus the formula is modified to:
\begin{equation}
\label{eq:simple_quantum_weight_update_rule}
{\bm{\theta}}_i^{t+1} = {\bm{\theta}}_i^t  + 2 \alpha^t h^t \frac{\partial}{\partial\bm{\theta}}K(\bm{x}, \bm{\theta})\Big|_{\bm{\theta}_i^t}.
\end{equation}
Similar to the original version of SOM, along with the learning rate $\alpha$, the neighborhood function is a hyperparameter in the training process; however, in our case, Gaussian functional form is chosen, as given in Eq.~(\ref{eq:neighborhood}).
The training process results in a map $\kappa(\rho)$ from the high-dimensional space $\mathcal{B(H)}$ to $L$, which preserves the topological structure present in $\mathcal{B(H)}$. 
In the inference phase, the weights are no longer updated, but simply the appropriate BMU $l^*$ for a data sample $\rho(\bm{x}  )$ is selected.
The gradients of the kernel functions, as specified in Eq.~(\ref{eq:simple_quantum_weight_update_rule}) can be calculated using the parameter shift rule~\cite{wierichs2022general} given by
\begin{equation}
\label{eq:parameter_shift}
\frac{\partial}{\partial \theta}K(x, \theta)\Big|_{\theta_i^t} = 
\frac{K(x, \theta +\phi) - K(x, \theta -\phi)}{2\sin(\omega \phi)/\omega},
\end{equation}
where $\omega$ corresponds to the eigenvalues of the Pauli gates.

\section{\label{sec:experiments} Numerical Experiments}
In order to establish the efficacy of the proposed variational quantum self-organizing map, we employ this algorithm to generate two distinct low-dimensional topology-preserving maps. 
One map is constructed using classical data, while the other map is constructed using quantum data. 
In the first scenario, the lower-dimensional map is employed for the purpose of clustering and for creating a low-dimensional topology-preserving representation of the three distinct species of flowers from the Iris data set~\cite{misc_iris_53}.
In the second scenario, the proposed algorithm is employed to generate a low-dimensional projection map of the state space of a $1+1$-dimensional lattice gauge theory describing the interaction between electrons and photons, called Schwinger model~\cite{PhysRev.128.2425}. 
The resultant low-dimensional map is employed to distinguish between two distinct phases of the model.

\subsection{\label{sub_sec:qsom_on_classical_data} Variational QSOM on Iris data}
The Iris dataset, developed by Ronald Fisher, is widely recognized and frequently employed within the machine learning field. 
The dataset comprises of $150$ samples and $4$ features. 
The dataset contains $50$ samples from each of the three Iris species, namely Iris setosa, Iris virginica, and Iris versicolor.
The four features provide data on the measurements of the sepals and petals, specifically their length and width, for each sample. 
The dataset also contains an appropriate label which corresponds to a particular type of the species of the Iris flower.
However, note that, we do not use this label during the training phase.
Our primary aim, in fact, is to evaluate the unsupervised learning capability of the algorithm which utilizes only the raw features of the data.
In the training phase the algorithm is trained to generate hidden clusters in the data set, and in the inference phase a test sample is assigned to an appropriate cluster. 
The resultant cluster label is then compared against the corresponding label for that sample. 

For the data encoding of the classical data vector and the weight vector into the quantum states, we utilize \textit{ZZFeatureMap} available in Qiskit~\cite{Qiskit}. 
We use the nearest neighbour (i.e. `pairwise') entanglement. 
The corresponding unitary is given as
\begin{equation}
U_{\Phi(\textbf{x})} = U(\textbf{x})H^{\otimes n},
\end{equation}  
where
\begin{equation}
\label{eq:Ising_mapping}
U(\textbf{x}) = \exp\Big(i\sum_i\phi(x_i)Z_i + i\sum_{<ij>}\phi'(x_i, x_{j})Z_iZ_{j}\Big).
\end{equation}
In principle, it is possible to customize the functions $\phi$ and $\phi'$; however, in our experiments, we use $\phi(x_i) = 2 x_i$ and
$\phi'(x_i, x_{i+1}) = 2(x_i-\pi)(x_{i+1}-\pi)$.
The input data is scaled from $-1$ to $1$.
It is conjectured that for such a map, the transition probability $|\langle\Psi(\bm{\theta}_i)|\Psi(\bm{x})\rangle|^2$ (i.e. the overlap between the input quantum state and the quantum state parametrized by the weight vector), which can be estimated on a quantum computer relatively easily, is hard to evaluate on a classical computer~\cite{bremner2011classical}. 

The output layer in our architecture consists of $36$ neurons arranged in a grid of size $(6, 6)$. 
The weight vectors on each of the nodes in the output layer are randomly initialized, with each component taking values within the range of $-\pi/2$ to $\pi/2$. 
The learning rate $\alpha(t)$ is initialized to $1$ and decreased with an exponential pre-factor (which depends on the number of training iterations) during the training process.
Similarly, the variance $\sigma$, which determines the extent of interaction between neighboring neurons, is initially set to $5$ and iteratively reduced with an iteration-dependent exponentially decreasing pre-factor. 
The algorithm is trained on $500$ randomly picked samples from the entire dataset. Note that this step necessitates the repetition of some of the samples. 
During the validation phase, the assignment of data samples to the relevant clusters is accomplished by identifying the matching BMU for each sample.

\begin{figure}[h!]
\begin{center}
\frame{\includegraphics[width=0.48\textwidth]{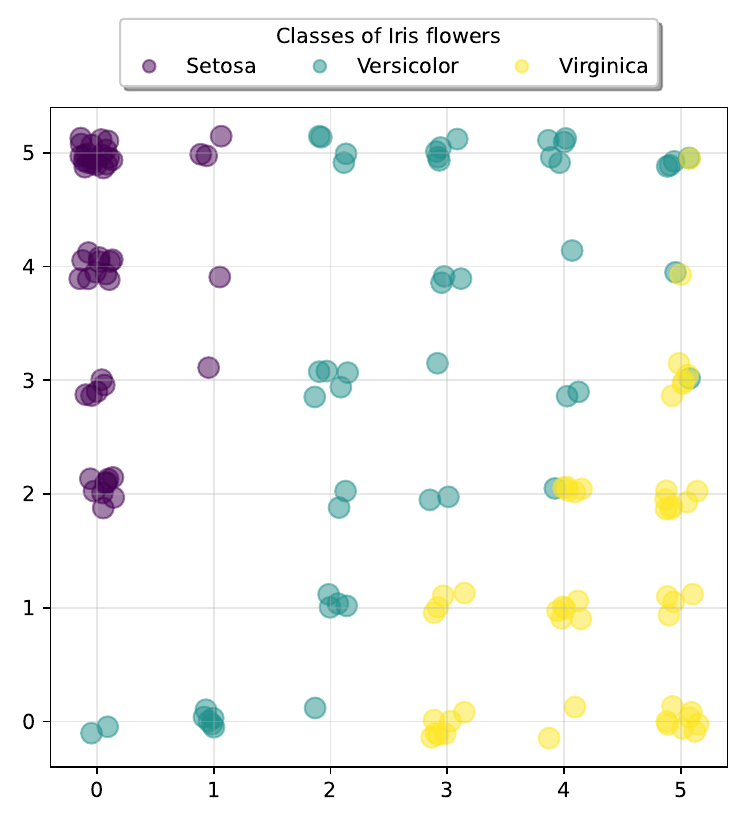}}
\caption{The variational quantum self-organizing map (variational QSOM) for the Iris dataset not only performs classification of distinct flowers, but also maintains the topological structure of the Hilbert space in the reduced-dimensional representation, that is the map maintains the topological characteristics of the original data by grouping together the quantum states from the same class.}
\label{fig:vqsom_iris_results}
\end{center}
\end{figure}

\begin{figure}
    \centering
    \begin{tikzpicture}
\node[anchor=north west] at (-3.5, 2.5) {\frame{\includegraphics[width=0.47\textwidth]{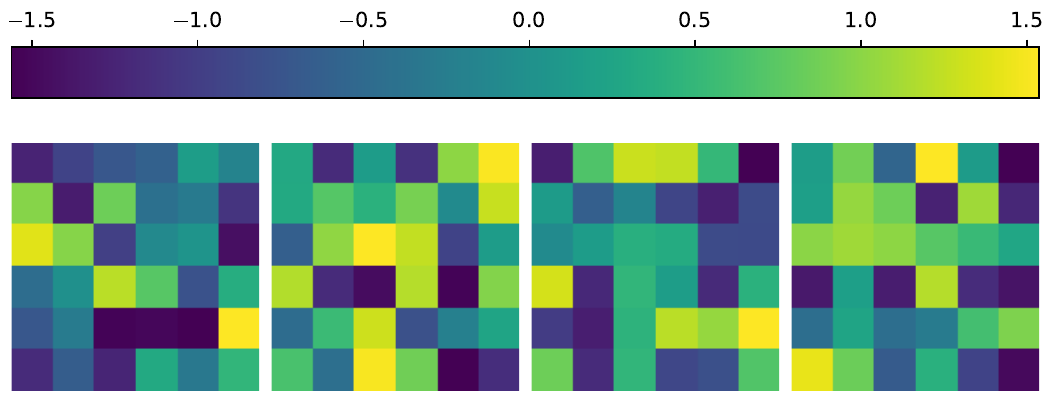}}};
\node[anchor=north west] at (-3.5, -2.0) {\frame{\includegraphics[width=0.47\textwidth]{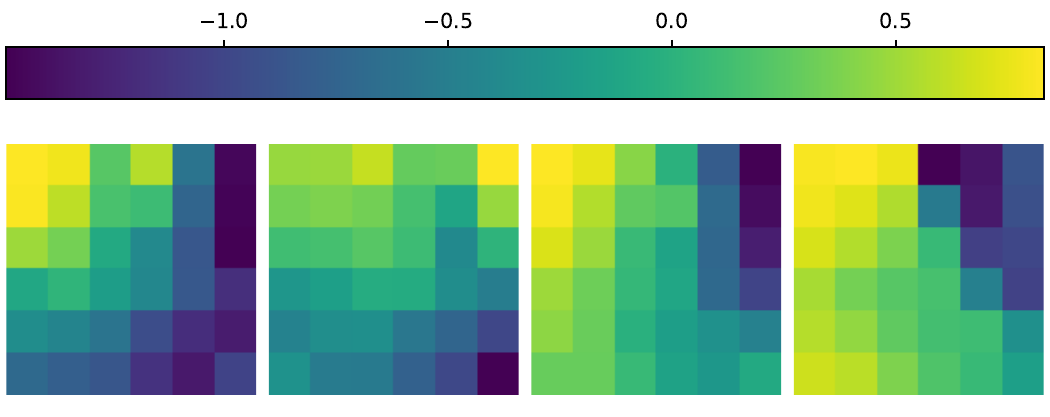}}};
\node at (0.8, -1.35) {(a)};
\node at (0.8, -5.85) {(b)};
\end{tikzpicture}
        \caption{
        (a) The numerical values of the four components of the weight vectors at the beginning of the training process. 
        (b) Corresponding numerical values at the end of the training process.
        }
     \label{fig:20_qubits}
\end{figure}

The resultant two-dimensional map from the validation phase is shown in Fig.~\ref{fig:vqsom_iris_results}. 
The map exhibits a clear capacity to differentiate and categorize the groupings of 3 different species of the Iris flower. 
Furthermore, the map effectively maintains the topological characteristics of the original data by grouping together samples with similar attributes on adjacent nodes in the output layer.
It is crucial to emphasize that the labels were not included as an input to the algorithm, but are instead provided here for the purpose of facilitating comparison.
In Fig.~\ref{fig:20_qubits}, we also display the numerical values of the four components of the weight vectors at the beginning of the training process and the corresponding numerical values at the end of the training process.
Such smoother patterns of the numerical values of the components are characteristic of the trained self-organizing maps.

\subsection{\label{sub_sec:qsom_on_quantum_data} Variational QSOM on lattice Schwinger model}

The Schwinger model plays a significant role in the field of high-energy physics due to its similarities to the quantum theory that describes the interactions between quarks and gluons, i.e. quantum chromodynamics. These similarities include phenomena such as spontaneous symmetry breaking, fermion confinement, charge shielding, and the presence of a topological $\theta$ vacuum. 
Furthermore, it serves as an ideal model for evaluating the effectiveness of quantum computers in examining the static and dynamical properties of lattice gauge theories~\cite{PhysRevA.90.042305, PhysRevX.3.041018, kokail2019self, PhysRevD.106.054508, PRXQuantum.3.020324, farrell2023scalable, savage2023quantum, angelides2023first}. 

The Schwinger model is an abelian gauge theory in $1+1$ dimensions, where the gauge group $U(1)$ governs the interaction between fermions and gauge bosons. 
It serves as a theoretical framework to illustrate the well-known Schwinger mechanism, a phenomenon characterized by the spontaneous generation of particle-antiparticle pairs in the presence of strong electric field~\cite{PhysRev.128.2425}. 
Coleman established that the massive Schwinger model displays a phase transition at $\theta = \pi$ occurring at a critical point determined by certain values of $m$ (fermion mass) and $g$ (coupling constant)~\cite{COLEMAN1976239}. 
In this study, we utilize the variational quantum self-organizing map to distinguish between the two distinct phases characterized by this phase transition. 

The problem of classifying the two distinct phases of the Schwinger model using quantum machine learning techniques has been previously investigated by K\"{u}hn et al.~\cite{PhysRevA.90.042305}. 
However, the primary distinction between our methodology and theirs is the implementation of an unsupervised machine learning strategy in our approach, which does not necessitate the presence of labels. 
Nevertheless, our framework for data generation closely follows the one described in~\cite{PhysRevA.90.042305}.
We describe the process of data generation below. 
The Lagrangian of the Schwinger model is given by:
\begin{equation}
\label{eq:schwinger_lagrangian}
\mathcal{L} = 
\psi(i\gamma^{\mu} D_{\mu} - m )\psi 
-\frac{1}{4}F_{\mu\nu}F^{\mu\nu}
+ \frac{g\theta}{4\pi}\epsilon_{\mu\nu}F^{\mu\nu},
\end{equation}
where $\psi$ corresponds to the fermion field,  $D_{\mu} = \partial_{\mu} + igA_{\mu}$ is the covariane derivative, and $A_{\mu}$ corresponds to the gauge field. The coupling constant $g$ governs the strength of the interation between the fermions and gauge bosons, $m$ is the fermion mass, and $F_{\mu\nu}$ corresponds to the electromagnetic tensor. 
The lattice implementation of the Schwinger model can be achieved by employing the staggered fermion techniques~\cite{PhysRevD.11.395}.
After the Jordan-Wigner transformation the Hamiltonian of the Schwinger model is given by
\begin{align}
\label{eq:schwinger_jordan_wigner}
H
&= J \sum_{n=0}^{N_s-2}\bigg(\sum_{i=0}^{n}\frac{Z+(-1)^i}{2} + \frac{\theta}{2\pi}\bigg)^2 \nonumber \\
&+ \frac{w}{2}\sum_{n=0}^{N_2-2}[X_nX_{n+1}+Y_nY_{n+1}] 
+ \frac{m}{2}\sum_{n=0}^{N_2-1}(-1)^n Z_n.
\end{align}

The Hamiltonian (with $N_s=4$, i.e. 4 spatial lattice sites) is diagonalized in order to derive the lowest eigenstates corresponding to a specific value of the ratio $m$/$g$. 
The eigenstates are then labeled based on the order parameter that governs the phase transition of the Schwinger model, i.e. the expectation value of an averaged electric field 
\begin{equation}
    E = \frac{1}{N}\sum_{n=0}^{N-1}\sum_{n=0}^{N-1}\frac{Z_i+ (-1)^i}{2}.
\end{equation}
If $\langle E\rangle=0$, then the corresponding state is labeled as class `$0$'. Conversely, if $\langle E\rangle>0$, it is denoted as the class `$1$'.
It is crucial to emphasize that, similar to the unsupervised learning of the Iris dataset previously described, the labels are not included as input to the algorithm; instead, they are provided for the purpose of comparison.

\begin{figure}
\begin{center}
\frame{\includegraphics[width=0.48\textwidth]{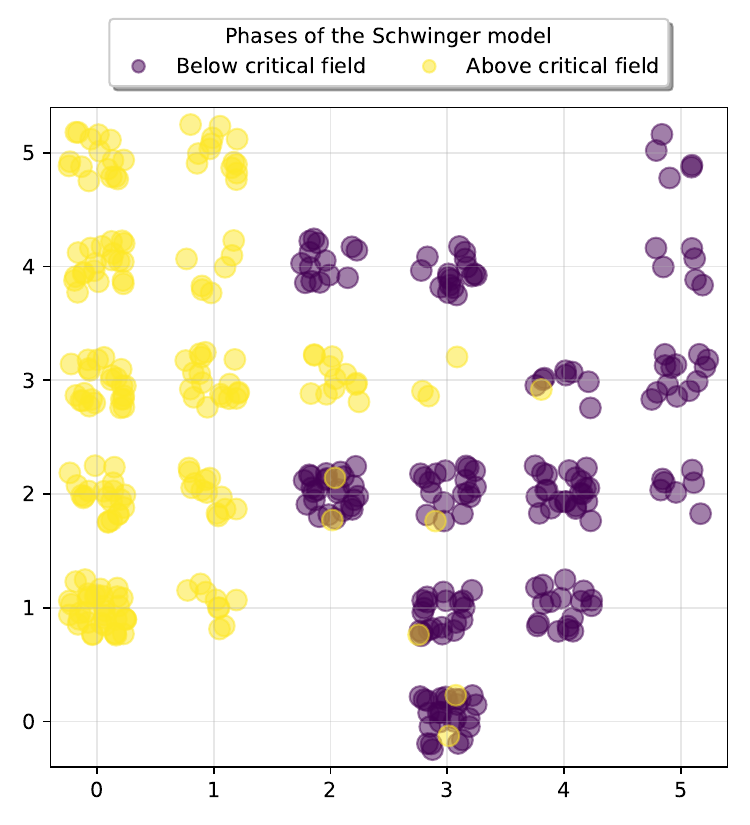}}
\caption{The variational quantum self-organizing map (variational QSOM) of the Schwinger model not only performs classification of distinct phases, but also maintains the topological structure of the Hilbert space in the reduced-dimensional representation, that is the map maintains the topological characteristics of the original data by grouping together the quantum states from the same phase. The legends `Below critical field' and `Above critial field' refer to states with $\langle E\rangle=0$ and $\langle E\rangle>0$, respectively.}
\label{fig:qom_map_schwinger}
\end{center}
\end{figure}

We observe that the variational quantum self-organizing map (variational QSOM) not only performs classification of distinct phases, but also maintains the topological structure of the Hilbert space in the reduced-dimensional representation, that is, the map maintains the topological characteristics of the original data by grouping together the quantum states from the same phase, as shown in Fig.~\ref{fig:qom_map_schwinger}. 

\section{\label{sec:conclusion}Discussion}

In this article, we have proposed a novel quantum neural network architecture for unsupervised machine learning of classical and quantum data on a near-term quantum computer.
The architecture underlying the proposed quantum neural network is inspired by and based on Kohonen's \textit{self-organizing map} (SOM).
The proposed algorithm scales linearly in the number of quantum data samples provided that there exists a predefined number of quantum states parametrized by the weight vectors in the output layer of SOM.
We have demonstrated the unsupervised learning capability of the proposed quantum algorithm by creating low-dimensional representation maps for both classical and quantum datasets. 
The algorithm generates lower-dimensional maps that preserve the topological properties by clustering quantum states of the same classes.
It is important to note that multiple steps of this algorithm can be readily parallelized, including the search for the best matching unit and the inference phase.

Several investigations remain for future exploration. 
The objective of this article was to elucidate the quantum algorithm; consequently, a comparison with the classical self-organizing map was not conducted.
A rigorous comparison with a classical technique can be conducted using many performance indicators, including Fowlkes-Mallows scores~\cite{Fowlkes01091983}, Silhouette Coefficient~\cite{ROUSSEEUW198753}, Davies-Bouldin Index~\cite{4766909}, and Calinski-Harabasz Index~\cite{Caliński01011974}. 
In terms of these scores, it would be insightful to compare the classical version of the self-organizing maps, as well as other classical clustering and dimensionality reduction techniques, to the variational quantum self-organizing map.
Additionally, the efficiency and effectiveness of our algorithm may be enhanced by utilizing projected quantum kernels instead of fidelity kernels, as they exhibit greater geometric differences than fidelity-based kernels~\cite{huang2021power}. 
An additional intriguing topic that merits further investigation is the relationship and contrast between the classical shadows formalism and our approach, along with its implications for supervised and unsupervised machine learning tasks~\cite{doi:10.1126/science.abk3333}.

The variational quantum self-organizing map's applicability may offer a better understanding of the distinct phases of quantum many-body systems. Nevertheless, it remains to be determined whether quantum kernels provide a competitive advantage over classical kernels in relation to classical datasets~\cite{ablan2025similarity}.
Lastly, the efficacy of the algorithm as the number of qubits increases and the impact of noise from its implementation on quantum hardware are topics that require further investigation.

\section{\label{sec:acknoledge}Acknowledgment}
Numerous discussions with Benjamin Boor and Nicolas Robles were pivotal in the preliminary stages of the algorithm's formulation.
The author expresses gratitude to Kunal Sharma, Shashanka Ubaru, Vladimir Rastunkov, Das Pemmaraju, Jae-Eun Park, Brian Quanz, Vaibhaw Kumar, and Daniel Fry for their constructive talks on these subjects. 

IBM, the IBM logo, and ibm.com are trademarks of International Business Machines Corp., registered in many jurisdictions worldwide. Other product and service names might be trademarks of IBM or other companies. The current list of IBM trademarks is available at \url{https://www.ibm.com/legal/copytrade}.
\bibliography{aps}

\end{document}